\documentclass[10pt,a4paper,onecolumn]{article}
\usepackage{marginnote}
\usepackage{graphicx}
\usepackage{xcolor}
\usepackage{authblk,etoolbox}
\usepackage{titlesec}
\usepackage{calc}
\usepackage{tikz}
\usepackage{hyperref}
\hypersetup{colorlinks,breaklinks=true,
            urlcolor=[rgb]{0.0, 0.5, 1.0},
            linkcolor=[rgb]{0.0, 0.5, 1.0}}
\usepackage{caption}
\usepackage{tcolorbox}
\usepackage{amssymb,amsmath}
\usepackage{ifxetex,ifluatex}
\usepackage{seqsplit}
\usepackage{xstring}

\usepackage{float}
\let\origfigure\figure
\let\endorigfigure\endfigure
\renewenvironment{figure}[1][2] {
    \expandafter\origfigure\expandafter[H]
} {
    \endorigfigure
}

\usepackage{fixltx2e} 

\let\textttOrig=\texttt
\def\texttt#1{\expandafter\textttOrig{\seqsplit{#1}}}
\renewcommand{\seqinsert}{\ifmmode
  \allowbreak
  \else\penalty6000\hspace{0pt plus 0.02em}\fi}

\makeatletter
\let\href@Orig=\href
\def\href@Urllike#1#2{\href@Orig{#1}{\begingroup
    \def\Url@String{#2}\Url@FormatString
    \endgroup}}
\def\href@Notdoi#1#2{\def\tempa{#1}\def\tempb{#2}%
  \ifx\tempa\tempb\relax\href@Urllike{#1}{#2}\else
  \href@Orig{#1}{#2}\fi}
\def\href#1#2{%
  \IfBeginWith{#1}{https://doi.org}%
  {\href@Urllike{#1}{#2}}{\href@Notdoi{#1}{#2}}}
\makeatother

\newlength{\cslhangindent}
\newlength{\csllabelwidth}
\newenvironment{CSLReferences}[3] 
 {
  \setlength{\parindent}{0pt}
  \ifodd #1 \everypar{\setlength{\hangindent}{\cslhangindent}}\ignorespaces\fi
  \ifnum #2 > 0
  \setlength{\parskip}{#2\baselineskip}
  \fi
 }%
 {}
\usepackage{calc}

\usepackage[top=3.5cm, bottom=3cm, right=1.5cm, left=1.0cm,
            headheight=2.2cm, reversemp, includemp, marginparwidth=4.5cm]{geometry}

\titleformat{\section}
  {\normalfont\sffamily\Large\bfseries}
  {}{0pt}{}
\titleformat{\subsection}
  {\normalfont\sffamily\large\bfseries}
  {}{0pt}{}
\titleformat{\subsubsection}
  {\normalfont\sffamily\bfseries}
  {}{0pt}{}
\titleformat*{\paragraph}
  {\sffamily\normalsize}

\usepackage{fancyhdr}
\makeatletter
\let\ps@plain\ps@fancy
\fancyheadoffset[L]{4.5cm}
\fancyfootoffset[L]{4.5cm}

\definecolor{linky}{rgb}{0.0, 0.5, 1.0}

\newtcolorbox{repobox}
   {colback=red, colframe=red!75!black,
     boxrule=0.5pt, arc=2pt, left=6pt, right=6pt, top=3pt, bottom=3pt}

\patchcmd{\@maketitle}{center}{flushleft}{}{}
\patchcmd{\@maketitle}{center}{flushleft}{}{}
\patchcmd{\@maketitle}{\LARGE}{\LARGE\sffamily}{}{}
\def\maketitle{{%
  
  \AB@maketitle}}
\makeatletter
\renewcommand\AB@affilsepx{ \protect\Affilfont}
\renewcommand\AB@affilnote[1]{{\bfseries #1}\hspace{3pt}}
\renewcommand{\affil}[2][]%
   {\newaffiltrue\let\AB@blk@and\AB@pand
      \if\relax#1\relax\def\AB@note{\AB@thenote}\else\def\AB@note{#1}%
        \setcounter{Maxaffil}{0}\fi
        \begingroup
        \let\href=\href@Orig
        \let\texttt=\textttOrig
        \let\protect\@unexpandable@protect
        \def\thanks{\protect\thanks}\def\footnote{\protect\footnote}%
        \@temptokena=\expandafter{\AB@authors}%
        {\def\\{\protect\\\protect\Affilfont}\xdef\AB@temp{#2}}%
         \xdef\AB@authors{\the\@temptokena\AB@las\AB@au@str
         \protect\\[\affilsep]\protect\Affilfont\AB@temp}%
         \gdef\AB@las{}\gdef\AB@au@str{}%
        {\def\\{, \ignorespaces}\xdef\AB@temp{#2}}%
        \@temptokena=\expandafter{\AB@affillist}%
        \xdef\AB@affillist{\the\@temptokena \AB@affilsep
          \AB@affilnote{\AB@note}\protect\Affilfont\AB@temp}%
      \endgroup
       \let\AB@affilsep\AB@affilsepx
}
\makeatother

\renewcommand\Affilfont{\sffamily\small\mdseries}
\ifnum 0\ifxetex 1\fi\ifluatex 1\fi=0 
  \usepackage[OT1]{fontenc}
  \usepackage[utf8]{inputenc}

\else 
  \ifxetex
    \usepackage{mathspec}
    \usepackage{fontspec}

  \else
    \usepackage{fontspec}
  \fi
  \defaultfontfeatures{Ligatures=TeX,Scale=MatchLowercase}

\fi
\IfFileExists{upquote.sty}{\usepackage{upquote}}{}
\IfFileExists{microtype.sty}{%
\usepackage{microtype}
\UseMicrotypeSet[protrusion]{basicmath} 
}{}

\usepackage{hyperref}
\hypersetup{unicode=true,
            pdftitle={PySDM v1: particle-based cloud modelling package for~warm-rain microphysics and aqueous chemistry},
            pdfborder={0 0 0},
            breaklinks=true}
\usepackage{color}
\usepackage{fancyvrb}

\DefineVerbatimEnvironment{Highlighting}{Verbatim}{commandchars=\\\{\}}
\newenvironment{Shaded}{}{}

\newcommand{\ControlFlowTok}[1]{\textcolor[rgb]{0.00,0.44,0.13}{\textbf{#1}}}

\newcommand{\DecValTok}[1]{\textcolor[rgb]{0.25,0.63,0.44}{#1}}

\newcommand{\FloatTok}[1]{\textcolor[rgb]{0.25,0.63,0.44}{#1}}

\newcommand{\ImportTok}[1]{#1}

\newcommand{\KeywordTok}[1]{\textcolor[rgb]{0.00,0.44,0.13}{\textbf{#1}}}
\newcommand{\NormalTok}[1]{#1}
\newcommand{\OperatorTok}[1]{\textcolor[rgb]{0.40,0.40,0.40}{#1}}

\newcommand{\SpecialCharTok}[1]{\textcolor[rgb]{0.25,0.44,0.63}{#1}}
\newcommand{\SpecialStringTok}[1]{\textcolor[rgb]{0.73,0.40,0.53}{#1}}
\newcommand{\StringTok}[1]{\textcolor[rgb]{0.25,0.44,0.63}{#1}}

\let\addcontentslineOrig=\addcontentsline
\def\addcontentsline#1#2#3{\bgroup
  \let\texttt=\textttOrig\addcontentslineOrig{#1}{#2}{#3}\egroup}
\let\markbothOrig\markboth
\def\markboth#1#2{\bgroup
  \let\texttt=\textttOrig\markbothOrig{#1}{#2}\egroup}
\let\markrightOrig\markright
\def\markright#1{\bgroup
  \let\texttt=\textttOrig\markrightOrig{#1}\egroup}

\usepackage{graphicx,grffile}
\makeatletter
\def\maxwidth{\ifdim\Gin@nat@width>\linewidth\linewidth\else\Gin@nat@width\fi}
\def\maxheight{\ifdim\Gin@nat@height>\textheight\textheight\else\Gin@nat@height\fi}
\makeatother
\setkeys{Gin}{width=\maxwidth,height=\maxheight,keepaspectratio}
\IfFileExists{parskip.sty}{%
\usepackage{parskip}
}{
\setlength{\parindent}{0pt}
\setlength{\parskip}{6pt plus 2pt minus 1pt}
}
\providecommand{\tightlist}{%
  \setlength{\itemsep}{0pt}\setlength{\parskip}{0pt}}
\ifx\paragraph\undefined\else
\let\oldparagraph\paragraph
\renewcommand{\paragraph}[1]{\oldparagraph{#1}\mbox{}}
\fi
\ifx\subparagraph\undefined\else
\let\oldsubparagraph\subparagraph
\renewcommand{\subparagraph}[1]{\oldsubparagraph{#1}\mbox{}}
\fi

\title{PySDM v1: particle-based cloud modelling package for~warm-rain
microphysics and aqueous chemistry}

        \author[1]{Piotr Bartman}
          \author[1]{Oleksii~Bulenok}
          \author[1]{Kamil~Górski}
          \author[2]{Anna~Jaruga}
          \author[1, 3]{Grzegorz~Łazarski}
          \author[4]{Michael~Olesik}
          \author[1]{Bartosz~Piasecki}
          \author[2]{Clare~E.~Singer}
          \author[1]{Aleksandra~Talar}
          \author[5, 1]{Sylwester~Arabas}
    
      \affil[1]{Faculty of Mathematics and Computer Science,
Jagiellonian University, Kraków,~Poland ~~~~~~~~~~~~~~~~~~~~~~~}
      \affil[2]{Department of Environmental Science and Engineering,
California Institute of Technology, Pasadena,~CA,~USA ~~~~~~~~~~~~~~~~~~~~~~~~~~~~~~~~~~~~~~~~~~~~~~~~~~~~~~~~~~~~~~~~~~~~~~~~~~~~~~~~~~~~~~~~~~~~~~~~~~~~~~~~~~~~}
      \affil[3]{Faculty~of~Chemistry, Jagiellonian University, Kraków,
Poland ~~~~~~~~~~~~~~~~~~~~~~~~~~~~~~~~~~~~~~~~~~~~~~~~~~~~~~~~~~~~~~~~~~~~~~~~~~~~~~~~~~~~~~~~~~~~~~~~~~~~~~~}
      \affil[4]{Faculty of Physics, Astronomy and Applied Computer
Science, Jagiellonian University, Kraków, Poland ~~~~~~~~~~~~~~~~~~~~~~~~~~~~~~~~~~~~~~~~~~~~~~~~~~~~~~~~~~~~~~~~~~~~~~~~~~~~~~~~~~~~~~~~~~~~~~~~~~~~~~~~~~~~~~~~~~~~~~~~~~~~~~~~~~~~~~~~}
      \affil[5]{University of Illinois at Urbana-Champaign, Urbana, IL,
USA}
  \date{\vspace{-2ex}}

\begin{document}
\maketitle

\hypertarget{introduction}{%
\section{Introduction}\label{introduction}}

\texttt{PySDM} is an open-source Python package for simulating the
dynamics of particles undergoing condensational and collisional growth,
interacting with a fluid flow and subject to chemical composition
changes. It is intended to serve as a building block for process-level
as well as computational-fluid-dynamics simulation systems involving
representation of a continuous phase (air) and a dispersed phase
(aerosol), with \texttt{PySDM} being responsible for representation of
the dispersed phase. As of the major version 1 (v1), the development has
been focused on atmospheric cloud physics applications, in particular on
modelling the dynamics of particles immersed in moist air using the
particle-based approach to represent the evolution of the size spectrum
of aerosol/cloud/rain particles. The particle-based approach contrasts
the more commonly used bulk and bin methods in which atmospheric
particles are segregated into multiple categories (aerosol, cloud, rain)
and their evolution is governed by deterministic dynamics solved on the
same Eulerian grid as the dynamics of the continuous phase.
Particle-based methods employ discrete computational (super) particles
for modelling the dispersed phase. Each super particle is associated
with a set of continuously-valued attributes evolving in Lagrangian
manner. Such approach is particularly well suited for using
probabilistic representation of particle collisional growth
(coagulation) and for representing processes dependent on numerous
particle attributes which helps to overcome the limitations of bulk and
bin methods (Morrison et al., 2020).

The \texttt{PySDM} package core is a Pythonic high-performance
implementation of the Super-Droplet Method (SDM) Monte-Carlo algorithm
for representing collisional growth (Shima et al., 2009), hence the
name. The SDM is a probabilistic alternative to the mean-field approach
embodied by the Smoluchowski equation, for a comparative outline of both
approaches see Bartman \& Arabas (2021). In atmospheric aerosol-cloud
interactions, particle collisional growth is responsible for formation
of rain drops through collisions of smaller cloud droplets (warm-rain
process) as well as for aerosol washout.

Besides collisional growth, \texttt{PySDM} includes representation of
condensation/evaporation of water vapour on/from the particles.
Furthermore, representation of dissolution and, if applicable,
dissociation of trace gases (sulfur dioxide, ozone, hydrogen peroxide,
carbon dioxide, nitric acid and ammonia) is included to model the
subsequent aqueous-phase oxidation of the dissolved sulfur dioxide.
Representation of the chemical processes follows the particle-based
formulation of Jaruga \& Pawlowska (2018).

The usage examples are built on top of four different
\texttt{environment} classes included in \texttt{PySDM} v1 and
implementing common simple atmospheric cloud modelling frameworks: box,
adiabatic parcel, single-column and 2D prescribed flow kinematic models.

In addition, the package ships with tutorial code depicting how
\texttt{PySDM} can be used from \texttt{Julia} and \texttt{Matlab} using
the \texttt{PyCall.jl} and the Matlab-bundled Python interface,
respectively. Two exporter classes are available as of time of writing
enabling storage of particle attributes in the VTK format and storage of
gridded products in netCDF format.

\hypertarget{dependencies-and-supported-platforms}{%
\section{Dependencies and supported
platforms}\label{dependencies-and-supported-platforms}}

PySDM essential dependencies are: \texttt{NumPy}, \texttt{SciPy},
\texttt{Numba}, \texttt{Pint} and \texttt{ChemPy} which are all free and
open-source software available via the PyPI platform. \texttt{PySDM}
ships with a setup.py file allowing installation using the \texttt{pip}
package manager (i.e.,
\texttt{pip\ install\ git+https://github.com/atmos-cloud-sim-uj/PySDM.git}).

\texttt{PySDM} has two alternative parallel number-crunching backends
available: multi-threaded CPU backend based on \texttt{Numba} (Lam et
al., 2015) and GPU-resident backend built on top of \texttt{ThrustRTC}
(Yang, 2020). The optional GPU backend relies on proprietary
vendor-specific CUDA technology, the accompanying non-free software and
drivers; \texttt{ThrustRTC} and \texttt{CURandRTC} packages are released
under the Anti-996 license.

The usage examples for \texttt{Python} were developed embracing the
\texttt{Jupyter} interactive platform allowing control of the
simulations via web browser. All Python examples are ready for use with
the \texttt{mybinder.org} and the \texttt{Google\ Colab} platforms.

Continuous integration infrastructure used in the development of PySDM
assures the targeted full usability on Linux, macOS and Windows
environments. Compatibility with Python versions 3.7 through 3.9 is
maintained as of time of writing. Test coverage for PySDM is reported
using the \texttt{codecov.io} platform. Coverage analysis of the backend
code requires execution with JIT-compilation disabled for the CPU
backend (e.g., using the \texttt{NUMBA\_DISABLE\_JIT=1} environment
variable setting). For the GPU backend, a purpose-built
\texttt{FakeThrust} class is shipped with \texttt{PySDM} which
implements a subset of the \texttt{ThrustRTC} API and translates C++
kernels into equivalent \texttt{Numba} parallel Python code for
debugging and coverage analysis.

The \texttt{Pint} dimensional analysis package is used for unit testing.
It allows asserting on the dimensionality of arithmetic expressions
representing physical formulae. In order to enable JIT compilation of
the formulae for simulation runs, a purpose-built
\texttt{FakeUnitRegistry} class that mocks the \texttt{Pint} API
reducing its functionality to SI prefix handling is used by default
outside of tests.

\hypertarget{api-in-brief}{%
\section{API in brief}\label{api-in-brief}}

In order to depict PySDM API with a practical example, the following
listings provide sample code roughly reproducing the Figure 2 from the
Shima et al. (2009) paper in which the SDM algorithm was introduced.

It is a coalescence-only set-up in which the initial particle size
spectrum is exponential and is deterministically sampled to match the
condition of each super-droplet having equal initial multiplicity, with
the multiplicity denoting the number of real particles represented by a
single computational particle referred to as a super-droplet:

\begin{Shaded}
\begin{Highlighting}[]
\ImportTok{from}\NormalTok{ PySDM.physics }\ImportTok{import}\NormalTok{ si}
\ImportTok{from}\NormalTok{ PySDM.initialisation.spectral\_sampling }\ImportTok{import}\NormalTok{ ConstantMultiplicity}
\ImportTok{from}\NormalTok{ PySDM.physics.spectra }\ImportTok{import}\NormalTok{ Exponential}

\NormalTok{n\_sd }\OperatorTok{=} \DecValTok{2} \OperatorTok{**} \DecValTok{17}
\NormalTok{initial\_spectrum }\OperatorTok{=}\NormalTok{ Exponential(}
\NormalTok{    norm\_factor}\OperatorTok{=}\FloatTok{8.39e12}\NormalTok{, scale}\OperatorTok{=}\FloatTok{1.19e5} \OperatorTok{*}\NormalTok{ si.um }\OperatorTok{**} \DecValTok{3}\NormalTok{)}
\NormalTok{attributes }\OperatorTok{=}\NormalTok{ \{\}}
\NormalTok{spectral\_sampling }\OperatorTok{=}\NormalTok{ ConstantMultiplicity(spectrum}\OperatorTok{=}\NormalTok{initial\_spectrum)}
\NormalTok{attributes[}\StringTok{\textquotesingle{}volume\textquotesingle{}}\NormalTok{], attributes[}\StringTok{\textquotesingle{}n\textquotesingle{}}\NormalTok{] }\OperatorTok{=}\NormalTok{ spectral\_sampling.sample(n\_sd}\OperatorTok{=}\NormalTok{n\_sd)}
\end{Highlighting}
\end{Shaded}

In the above snippet, the \texttt{si} is an instance of the
\texttt{FakeUnitRegistry} class. The exponential distribution of
particle volumes is sampled at \(2^{17}\) points in order to initialise
two key attributes of the super-droplets, namely their volume and
multiplicity. Subsequently, a \texttt{Builder} object is created to
orchestrate dependency injection while instantiating the
\texttt{Particulator} class of \texttt{PySDM}:

\begin{Shaded}
\begin{Highlighting}[]
\ImportTok{import}\NormalTok{ numpy }\ImportTok{as}\NormalTok{ np}
\ImportTok{from}\NormalTok{ PySDM.builder }\ImportTok{import}\NormalTok{ Builder}
\ImportTok{from}\NormalTok{ PySDM.environments }\ImportTok{import}\NormalTok{ Box}
\ImportTok{from}\NormalTok{ PySDM.dynamics }\ImportTok{import}\NormalTok{ Coalescence}
\ImportTok{from}\NormalTok{ PySDM.physics.coalescence\_kernels }\ImportTok{import}\NormalTok{ Golovin}
\ImportTok{from}\NormalTok{ PySDM.backends }\ImportTok{import}\NormalTok{ CPU}
\ImportTok{from}\NormalTok{ PySDM.products }\ImportTok{import}\NormalTok{ ParticlesVolumeSpectrum}

\NormalTok{radius\_bins\_edges }\OperatorTok{=}\NormalTok{ np.logspace(}
\NormalTok{    np.log10(}\DecValTok{10} \OperatorTok{*}\NormalTok{ si.um), np.log10(}\FloatTok{5e3} \OperatorTok{*}\NormalTok{ si.um), num}\OperatorTok{=}\DecValTok{32}\NormalTok{)}

\NormalTok{builder }\OperatorTok{=}\NormalTok{ Builder(n\_sd}\OperatorTok{=}\NormalTok{n\_sd, backend}\OperatorTok{=}\NormalTok{CPU())}
\NormalTok{builder.set\_environment(Box(dt}\OperatorTok{=}\DecValTok{1} \OperatorTok{*}\NormalTok{ si.s, dv}\OperatorTok{=}\FloatTok{1e6} \OperatorTok{*}\NormalTok{ si.m }\OperatorTok{**} \DecValTok{3}\NormalTok{))}
\NormalTok{builder.add\_dynamic(Coalescence(kernel}\OperatorTok{=}\NormalTok{Golovin(b}\OperatorTok{=}\FloatTok{1.5e3} \OperatorTok{/}\NormalTok{ si.s)))}
\NormalTok{products }\OperatorTok{=}\NormalTok{ [ParticlesVolumeSpectrum(radius\_bins\_edges)]}
\NormalTok{particulator }\OperatorTok{=}\NormalTok{ builder.build(attributes, products)}
\end{Highlighting}
\end{Shaded}

\begin{figure}[h]
    \centering
    \includegraphics[width=0.7\textwidth]{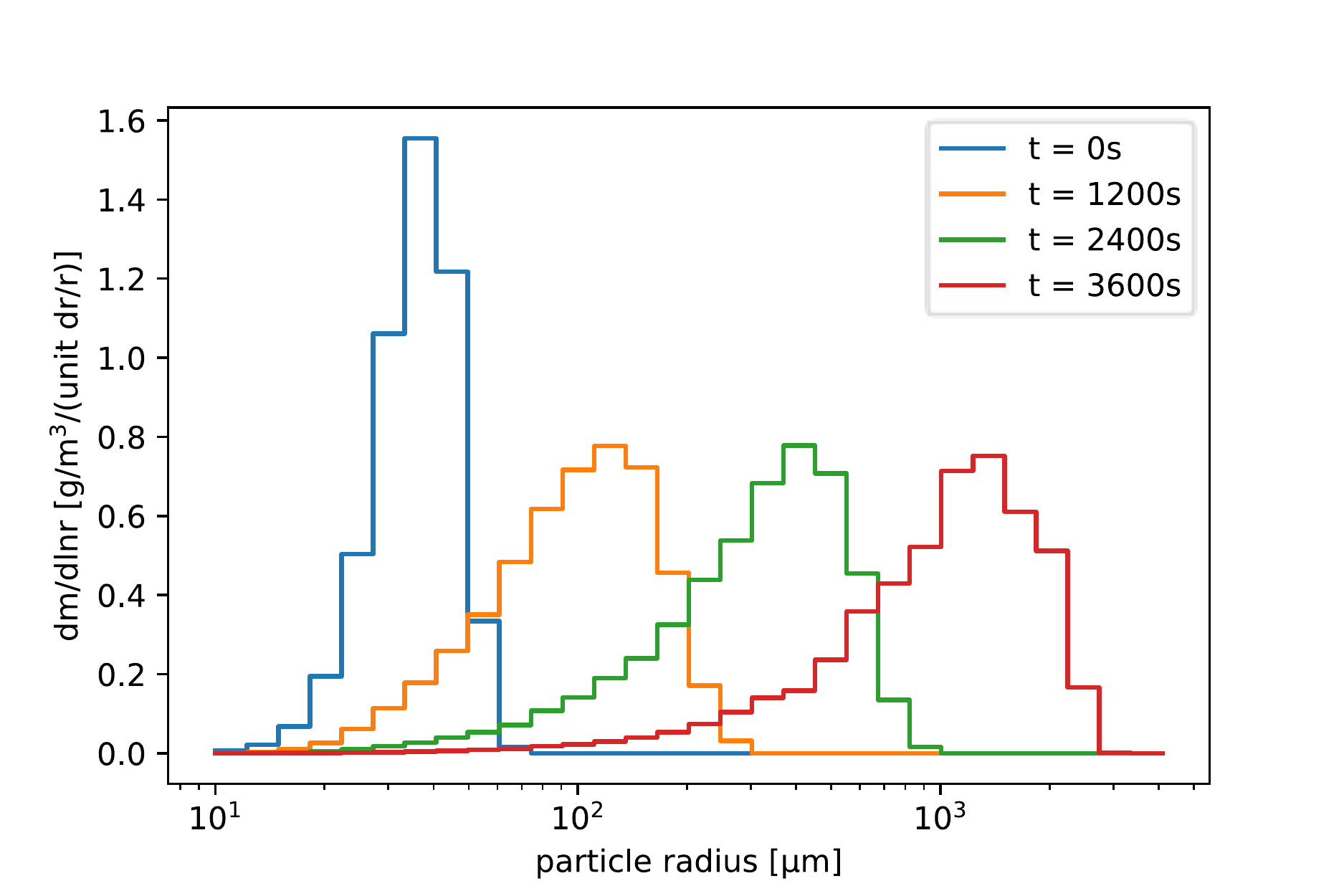}
    \caption{Sample plot generated with the code snippets included in the paper.}
    \label{fig:readme_fig_1}
\end{figure}

The \texttt{backend} argument may be set to an instance of either
\texttt{CPU} or \texttt{GPU} what translates to choosing the
multi-threaded \texttt{Numba}-based backend or the
\texttt{ThrustRTC-based} GPU-resident computation mode, respectively.
The employed \texttt{Box} environment corresponds to a zero-dimensional
framework (particle positions are neglected). The SDM Monte-Carlo
coalescence algorithm is added as the only dynamic in the system (other
dynamics available as of v1.3 represent condensational growth, particle
displacement, aqueous chemistry, ambient thermodynamics and Eulerian
advection). Finally, the \texttt{build()} method is used to obtain an
instance of the \texttt{Particulator} class which can then be used to
control time-stepping and access simulation state through the products
registered with the builder. A minimal simulation example is depicted
below with a code snippet and a resultant plot
(\autoref{fig:readme_fig_1}):

\begin{Shaded}
\begin{Highlighting}[]
\ImportTok{from}\NormalTok{ PySDM.physics.constants }\ImportTok{import}\NormalTok{ rho\_w}
\ImportTok{from}\NormalTok{ matplotlib }\ImportTok{import}\NormalTok{ pyplot}

\ControlFlowTok{for}\NormalTok{ step }\KeywordTok{in}\NormalTok{ [}\DecValTok{0}\NormalTok{, }\DecValTok{1200}\NormalTok{, }\DecValTok{2400}\NormalTok{, }\DecValTok{3600}\NormalTok{]:}
\NormalTok{    particulator.run(step }\OperatorTok{{-}}\NormalTok{ particulator.n\_steps)}
\NormalTok{    pyplot.step(}
\NormalTok{        x}\OperatorTok{=}\NormalTok{radius\_bins\_edges[:}\OperatorTok{{-}}\DecValTok{1}\NormalTok{] }\OperatorTok{/}\NormalTok{ si.um,}
\NormalTok{        y}\OperatorTok{=}\NormalTok{particulator.products[}\StringTok{\textquotesingle{}dv/dlnr\textquotesingle{}}\NormalTok{].get()[}\DecValTok{0}\NormalTok{] }\OperatorTok{*}\NormalTok{ rho\_w}\OperatorTok{/}\NormalTok{si.g,}
\NormalTok{        where}\OperatorTok{=}\StringTok{\textquotesingle{}post\textquotesingle{}}\NormalTok{, label}\OperatorTok{=}\SpecialStringTok{f"t = }\SpecialCharTok{\{}\NormalTok{step}\SpecialCharTok{\}}\SpecialStringTok{s"}\NormalTok{)}

\NormalTok{pyplot.xscale(}\StringTok{\textquotesingle{}log\textquotesingle{}}\NormalTok{)}
\NormalTok{pyplot.xlabel(}\StringTok{\textquotesingle{}particle radius [$\textbackslash{}mu$ m]\textquotesingle{}}\NormalTok{)}
\NormalTok{pyplot.ylabel(}\StringTok{"dm/dlnr [g/m$\^{}3$/(unit dr/r)]"}\NormalTok{)}
\NormalTok{pyplot.legend()}
\NormalTok{pyplot.show()}
\end{Highlighting}
\end{Shaded}

\hypertarget{usage-examples}{%
\section{Usage examples}\label{usage-examples}}

The PySDM examples are shipped in a separate package that can be
installed with \texttt{pip}
(\texttt{pip\ install\ git+https://github.com/atmos-cloud-sim-uj/PySDM-examples.git})
or conveniently experimented with using Colab or mybinder.org platforms
(single-click launching badges included in the \texttt{PySDM} README
file). The examples are based on setups from literature, and the package
is structured using bibliographic labels (e.g.,
\texttt{PySDM\_examples.Shima\_et\_al\_2009}).

All examples feature a \texttt{settings.py} file with simulation
parameters, a \texttt{simulation.py} file including logic analogous to
the one presented in the code snippets above for handling composition of
\texttt{PySDM} components using the \texttt{Builder} class, and a
Jupyter notebook file with simulation launching code and basic result
visualisation.

\hypertarget{box-environment-examples}{%
\subsubsection{Box environment
examples}\label{box-environment-examples}}

The \texttt{Box} environment is the simplest one available in PySDM and
the \texttt{PySDM\_examples} package ships with two examples based on
it. The first, is an extension of the code presented in the snippets in
the preceding section and reproduces Fig. 2 from the seminal paper of
Shima et al. (2009). Coalescence is the only process considered, and the
probabilities of collisions of particles are evaluated using the Golovin
additive kernel, which allows to compare the results with analytical
solution of the Smoluchowski equation (included in the resultant plots).

The second example based on the \texttt{Box} environment, also featuring
collision-only setup reproduces several figures from the work of Berry
(1966) involving more sophisticated collision kernels representing such
phenomena as the geometric sweep-out and the influence of electric field
on the collision probability.

\hypertarget{adiabatic-parcel-examples}{%
\subsubsection{Adiabatic parcel
examples}\label{adiabatic-parcel-examples}}

The \texttt{Parcel} environment shares the zero-dimensionality of
\texttt{Box} (i.e., no particle physical coordinates considered), yet
provides a thermodynamic evolution of the ambient air mimicking
adiabatic displacement of an air parcel in hydrostatically stratified
atmosphere. Adiabatic cooling during the ascent results in reaching
supersaturation what triggers activation of aerosol particles
(condensation nuclei) into cloud droplets through condensation. All
examples based on the \texttt{Parcel} environment utilise the
\texttt{Condensation} and \texttt{AmbientThermodynamics} dynamics.

The simplest example uses a monodisperse particle spectrum represented
with a single super-droplet and reproduces simulations described in
Arabas \& Shima (2017) where an ascent-descent scenario is employed to
depict hysteretic behaviour of the activation/deactivation phenomena.

A polydisperse lognormal spectrum represented with multiple
super-droplets is used in the example based on the work of Yang et al.
(2018). Presented simulations involve repeated ascent-descent cycles and
depict the evolution of partitioning between activated and unactivated
particles. Similarly, polydisperse lognormal spectra are used in the
example based on Lowe et al. (2019), where additionally each lognormal
mode has a different hygroscopicity. The Lowe et al. (2019) example
additionally features representation of droplet surface tension
reduction by organics.

Finally, there are two examples featuring adiabatic parcel simulations
involving representation of the dynamics of chemical composition of both
ambient air and the droplet-dissolved substances, in particular focusing
on the oxidation of aqueous-phase sulfur. The examples reproduce the
simulations discussed in Kreidenweis et al. (2003) and in Jaruga \&
Pawlowska (2018).

\hypertarget{kinematic-prescribed-flow-examples}{%
\subsubsection{Kinematic (prescribed-flow)
examples}\label{kinematic-prescribed-flow-examples}}

Coupling of \texttt{PySDM} with fluid-flow simulation is depicted with
both 1D and 2D prescribed-flow simulations, both dependent on the
\texttt{PyMPDATA} package (Bartman et al., 2021) implementing the MPDATA
advection algorithm. For a review on MPDATA, see e.g., Smolarkiewicz
(2006).

Usage of the \texttt{kinematic\_1d} environment is depicted in an
example based on the work of Shipway \& Hill (2012), while the
\texttt{kinematic\_2d} environment is showcased with a Jupyter notebook
featuring an interactive user interface and allowing studying
aerosol-cloud interactions in drizzling stratocumulus setup based on the
work of Arabas et al. (2015).

\begin{figure}
\centering
\includegraphics{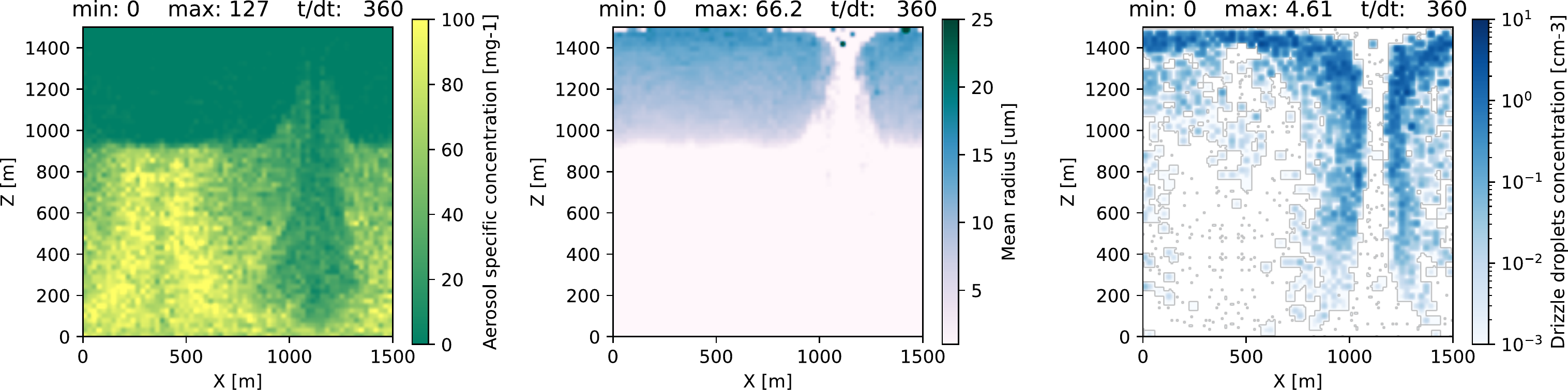}
\caption{Results from a 2D prescribed-flow simulation using the Arabas
et al. (2015) example.\label{fig:virga}}
\end{figure}

\autoref{fig:virga} presents a snapshot from the 2D simulation described
in detail in Arabas et al. (2015) and works cited therein. Each plot
depicts a 1.5 km by 1.5 km vertical slab of an idealised atmosphere in
which a prescribed single-eddy non-divergent flow is forced (updraft in
the left-hand part of the domain, downdraft in the right-hand part). The
left plot shows the distribution of aerosol particles in the air. The
upper part of the domain is covered with a stratocumulus-like cloud
which formed on the aerosol particles above the flat cloud base at the
level where relative humidity goes above 100\%. Within the cloud, the
aerosol concentration is thus reduced. The middle plot depicts the sizes
of particles. Particles larger than 1 micrometre in diameter are
considered as cloud droplets, particles larger than 50 micrometres in
diameter are considered as drizzle (unlike in bin or bulk models, such
categorisation is employed for analysis only and not within the
particle-based model formulation). Concentration of drizzle particles
forming through collisions is depicted in the right panel. A rain shaft
forms in the right part of the domain where the downward flow direction
amplifies particle sedimentation. Precipitating drizzle drops collide
with aerosol particles washing out the sub-cloud aerosol. Most of the
drizzle drops evaporate before reaching the bottom of the domain
depicting the virga phenomenon and the resultant aerosol resuspension.

\hypertarget{selected-relevant-recent-open-source-developments}{%
\section{Selected relevant recent open-source
developments}\label{selected-relevant-recent-open-source-developments}}

The SDM algorithm implementations are part of the following open-source
packages (of otherwise largely differing functionality):

\begin{itemize}
\tightlist
\item
  \texttt{libcloudph++} in C++ (Arabas et al., 2015; Jaruga \&
  Pawlowska, 2018) with Python bindings (Jarecka et al., 2015);
\item
  \texttt{SCALE-SDM} in Fortran, (Sato et al., 2018);
\item
  \texttt{PALM\ LES} in Fortran, (Maronga et al., 2020);
\item
  \texttt{LCM1D} in Python/C, (Unterstrasser et al., 2020);
\item
  \texttt{Pencil\ Code} in Fortran, (Brandenburg et al., 2021);
\item
  \texttt{NTLP} in Fortran, (Richter et al., 2021).
\item
  \texttt{superdroplet} in Python (\texttt{Cython} and \texttt{Numba}),
  C++, Fortran and Julia\\
  (\url{https://github.com/darothen/superdroplet});
\end{itemize}

List of links directing to SDM-related files within the above projects'
repositories is included in the \texttt{PySDM} README file.

Python packages for solving the dynamics of aerosol particles with
discrete-particle (moving-sectional) representation of the size spectrum
include (both depend on the \texttt{Assimulo} package for solving ODEs):

\begin{itemize}
\tightlist
\item
  \texttt{pyrcel}, (Rothenberg \& Wang, 2017);
\item
  \texttt{PyBox}, (Topping et al., 2018).
\end{itemize}

\hypertarget{summary}{%
\section{Summary}\label{summary}}

The key goal of the reported endeavour was to equip the cloud modelling
community with a solution enabling rapid development and
paper-review-level reproducibility of simulations (i.e., technically
feasible without contacting the authors and possible to be set up within
minutes) while being free from the two-language barrier commonly
separating prototype and high-performance research code. The key
advantages of PySDM stem from the characteristics of the employed Python
language which enables high performance computational modelling without
trading off such features as:

\begin{description}
    \item[succinct syntax]{ -- the snippets presented in the paper are arguably close to pseudo-code;}
    \item[portability]{depicted in PySDM with continuous integration Linux, macOS and Windows};
    \item[interoperability]{depicted in PySDM with Matlab and Julia usage examples requireing minimal amount of biding-specific code;}
    \item[multifaceted ecosystem]{depicted in PySDM with one-click execution of Jupyter notebooks on mybinder.org and colab.research.google.com platforms};
    \item[availability of tools for modern hardware]{depicted in PySDM with the GPU backend}.
\end{description}

PySDM together with a set of developed usage examples constitutes a tool
for research on cloud microphysical processes, and for testing and
development of novel modelling methods. PySDM is released under the GNU
GPL v3 license.

\hypertarget{author-contributions}{%
\section{Author contributions}\label{author-contributions}}

PB had been the architect and lead developer of PySDM v1 with SA taking
the role of main developer and maintainer over the time. PySDM 1.0
release accompanied PB's MSc thesis prepared under the mentorship of SA.
MO contributed to the development of the condensation solver and led the
development of relevant examples. GŁ contributed the initial draft of
the aqueous-chemistry extension which was refactored and incorporated
into PySDM under guidance from AJ. KG and BP contributed to the GPU
backend. CS and AT contributed to the examples. OB contributed the VTK
exporter. The paper was composed by SA and PB and is partially based on
the content of the PySDM README file and PB's MSc thesis.

\hypertarget{acknowledgements}{%
\section{Acknowledgements}\label{acknowledgements}}

We thank Shin-ichiro Shima (University of Hyogo, Japan) for his
continuous help and support in implementing SDM. We thank Fei Yang
(https://github.com/fynv/) for creating and supporting ThrustRTC.
Development of PySDM has been carried out within the
POWROTY/REINTEGRATION programme of the Foundation for Polish Science
co-financed by the European Union under the European Regional
Development Fund (POIR.04.04.00-00-5E1C/18).

\hypertarget{references}{%
\section*{References}\label{references}}
\addcontentsline{toc}{section}{References}

\hypertarget{refs}{}
\begin{CSLReferences}{1}{0}
\leavevmode\hypertarget{ref-Arabas_et_al_2015}{}%
Arabas, S., Jaruga, A., Pawlowska, H., \& Grabowski, W. W. (2015).
{l}ibcloudph++ 1.0: A single-moment bulk, double-moment bulk, and
particle-based warm-rain microphysics library in {C}++. \emph{Geosci.
Model Dev.} \url{https://doi.org/10.5194/gmd-8-1677-2015}

\leavevmode\hypertarget{ref-Arabas_and_Shima_2017}{}%
Arabas, S., \& Shima, S. (2017). On the CCN
(de)activation~nonlinearities. \emph{Nonlin. Process. Geophys.}
\url{https://doi.org/10.5194/npg-24-535-2017}

\leavevmode\hypertarget{ref-Bartman_and_Arabas_2021}{}%
Bartman, P., \& Arabas, S. (2021). On the design of {M}onte-{C}arlo
particle coagulation solver interface: A CPU/GPU super-droplet method
case study with PySDM. \emph{ArXiv e-Prints}.
\url{http://arxiv.org/abs/2101.06318}

\leavevmode\hypertarget{ref-Bartman_et_al_2021}{}%
Bartman, P., Banaśkiewicz, J., Drenda, S., Manna, M., Olesik, M.,
Rozwoda, P., Sadowski, M., \& Arabas, S. (2021). \emph{PyMPDATA v1:
Numba-accelerated implementation of MPDATA with examples in {P}ython,
{J}ulia and {M}atlab}. \url{https://pypi.org/p/PyMPDATA}

\leavevmode\hypertarget{ref-Berry_1966}{}%
Berry, E. X. (1966). Cloud droplet growth by collection. \emph{J. Atmos.
Sci.} \url{https://doi.org/1520-0469(1967)024\%3C0688:CDGBC\%3E2.0.CO;2}

\leavevmode\hypertarget{ref-Pencil_2021}{}%
Brandenburg, A., Johansen, A., Bourdin, P. A., Dobler, W., Lyra, W.,
Rheinhardt, M., Bingert, S., Haugen, N. E. L., Mee, A., Gent, F.,
Babkovskaia, N., Yang, C.-C., Heinemann, T., Dintrans, B., Mitra, D.,
Candelaresi, S., Warnecke, J., Käpylä, P. J., Schreiber, A., \ldots{}
Qian, C. (2021). The pencil code, a modular MPI code for partial
differential equations and particles: Multipurpose and
multiuser-maintained. \emph{J. Open Source Soft.}
\url{https://doi.org/10.21105/joss.02807}

\leavevmode\hypertarget{ref-Jarecka_et_al_2015}{}%
Jarecka, D., Arabas, S., \& Del Vento, D. (2015). Python bindings for
libcloudph++. \emph{ArXiv e-Prints}.
\url{http://arxiv.org/abs/1504.01161}

\leavevmode\hypertarget{ref-Jaruga_and_Pawlowska_2018}{}%
Jaruga, A., \& Pawlowska, H. (2018). {l}ibcloudph++ 2.0: Aqueous-phase
chemistry extension of the particle-based cloud microphysics scheme.
\emph{Geosci. Model Dev.} \url{https://doi.org/10.5194/gmd-11-3623-2018}

\leavevmode\hypertarget{ref-Kreidenweis_et_al_2003}{}%
Kreidenweis, S. M., Walcek, C. J., Feingold, G., Gong, W., Jacobson, M.
Z., Kim, C. H., Liu, X., Penner, J. E., Nenes, A., \& Seinfeld, J. H.
(2003). Modification of aerosol mass and size distribution due to
aqueous‐phase SO\(_2\) oxidation in clouds: Comparisons of several
models. \emph{J. Geophys. Res.}
\url{https://doi.org/10.1029/2002JD002673}

\leavevmode\hypertarget{ref-Numba}{}%
Lam, S. K., Pitrou, A., \& Seibert, S. (2015). Numba: A LLVM-based
python JIT compiler. \emph{Proceedings of the Second Workshop on the
LLVM Compiler Infrastructure in HPC}.
\url{https://doi.org/10.1145/2833157.2833162}

\leavevmode\hypertarget{ref-Lowe_et_al_2019}{}%
Lowe, S. J., Partridge, D. G., Davies, J. F., Wilson, K. R., Topping,
D., \& Riipinen, I. (2019). Key drivers of cloud response to
surface-active organics. \emph{Nature Comm.}
\url{https://doi.org/10.1038/s41467-019-12982-0}

\leavevmode\hypertarget{ref-Maronga_et_al_2020}{}%
Maronga, B., Banzhaf, S., Burmeister, C., Esch, T., Forkel, R.,
Fröhlich, D., Fuka, V., Gehrke, K., Geletič, J., Giersch, S.,
Gronemeier, T., Groß, G., Heldens, W., Hellsten, A., Hoffmann, F.,
Inagaki, A., Kadasch, E., Kanani-Sühring, F., Ketelsen, K., \& Raasch,
S. (2020). Overview of the {PALM} model system 6.0. \emph{Geosci. Model
Dev.} \url{https://doi.org/10.5194/gmd-13-1335-2020}

\leavevmode\hypertarget{ref-Morrison_et_al_2020}{}%
Morrison, H., Lier-Walqui, M. van, Fridlind, A. M., Grabowski, W. W.,
Harrington, J. Y., Hoose, C., Korolev, A., Kumjian, M. R., Milbrandt, J.
A., Pawlowska, H., Posselt, D. J., Prat, O. P., Reimel, K. J., Shima,
S., Diedenhoven, B. van, \& Xue, L. (2020). Confronting the challenge of
modeling cloud and precipitation microphysics. \emph{J. Adv. Model.
Earth Syst.} \url{https://doi.org/10.1029/2019MS001689}

\leavevmode\hypertarget{ref-Richter_et_al_2021}{}%
Richter, D. H., MacMillan, T., \& Wainwright, C. (2021). A {L}agrangian
cloud model for the study of marine fog. \emph{Boundary-Layer Meteorol.}
\url{https://doi.org/10.1007/s10546-020-00595-w}

\leavevmode\hypertarget{ref-Rothenberg_and_Wang_2017}{}%
Rothenberg, D., \& Wang, C. (2017). An aerosol activation metamodel of
v1.2.0 of the pyrcel cloud parcel model: Development and offline
assessment for use in an aerosol--climate model. \emph{Geosci. Model.
Dev.} \url{https://doi.org/10.5194/gmd-10-1817-2017}

\leavevmode\hypertarget{ref-Sato_et_al_2018}{}%
Sato, Y., Shima, S., \& Tomita, H. (2018). Numerical convergence of
shallow convection cloud field simulations: Comparison between
double‐moment {E}ulerian and particle‐based {L}agrangian microphysics
coupled to the same dynamical core. \emph{J. Adv. Model. Earth Syst.}
\url{https://doi.org/10.1029/2018MS001285}

\leavevmode\hypertarget{ref-Shima_et_al_2009}{}%
Shima, S., Kusano, K., Kawano, A., Sugiyama, T., \& Kawahara, S. (2009).
The super‐droplet method for the numerical simulation of clouds and
precipitation: A particle‐based and probabilistic microphysics model
coupled with a non‐hydrostatic model. \emph{Q. J. Royal Meteorol. Soc.}
\url{https://doi.org/10.1002/qj.441}

\leavevmode\hypertarget{ref-Shipway_and_Hill_2012}{}%
Shipway, B. J., \& Hill, A. A. (2012). Diagnosis of systematic
differences between multiple parametrizations of warm rain microphysics
using a kinematic framework. \emph{Q. J. Royal Meteorol. Soc.}
\url{https://doi.org/10.1002/qj.1913}

\leavevmode\hypertarget{ref-Smolarkiewicz_2006}{}%
Smolarkiewicz, P. K. (2006). Multidimensional positive definite
advection transport algorithm: An overview. \emph{Int. J. Numer. Methods
Fluids}. \url{https://doi.org/doi:10.1002/fld.1071}

\leavevmode\hypertarget{ref-Topping_et_al_2018}{}%
Topping, D., Connolly, P., \& Reid, J. (2018). {PyBox}: An automated
box-model generator for atmospheric chemistry and aerosol simulations.
\emph{J. Open Source Soft.} \url{https://doi.org/10.21105/joss.00755}

\leavevmode\hypertarget{ref-Unterstrasser_et_al_2020}{}%
Unterstrasser, S., Hoffmann, F., \& Lerch, M. (2020). Collisional growth
in a particle-based cloud microphysical model: Insights from column
model simulations using {LCM1D} (v1.0). \emph{Geosci. Model Dev.}
\url{https://doi.org/10.5194/gmd-13-5119-2020}

\leavevmode\hypertarget{ref-ThrustRTC}{}%
Yang, F. (2020). ThrustRTC: CUDA tool set for non-{C}++ languages that
provides similar functionality like {T}hrust, with NVRTC at its core. In
\emph{GitHub repository}. GitHub.
\url{https://github.com/fynv/thrustrtc}

\leavevmode\hypertarget{ref-Yang_et_al_2018}{}%
Yang, F., Kollias, P., Shaw, R. A., \& Vogelmann, A. M. (2018). Cloud
droplet size distribution broadening during diffusional growth: Ripening
amplified by deactivation and reactivation. \emph{Atmos. Chem. Phys.}
\url{https://doi.org/10.5194/acp-18-7313-2018}

\end{CSLReferences}

\end{document}